\documentclass[11pt,twoside]{article}
\usepackage{amsmath,amssymb,euscript,theorem}

\usepackage{setspace}

\newcommand{\e}{\varepsilon}

\newcommand{\al}{\alpha}

\newcommand{\pd}{\partial}

\newcommand{\R}{\mathbb{R}}
\newcommand{\Complex}{\mathbb{C}}
\renewcommand{\Re}{\mathop{\mathrm{Re}}}
\renewcommand{\Im}{\mathop{\mathrm{Im}}}
\newcommand{\mr}[1]{\mathrm{#1}}

\newcommand{\donothing}[1]{}
\newcommand{\myproof}{{\bf Proof.} \ }
\newcommand{\QED}{\quad $\square$} 
\newcommand{\Hc}{\EuScript{H}_c}

\numberwithin{equation}{section}
\newtheorem{theorem}{Theorem}[section]

\newtheorem{lemma}[theorem]{Lemma}

{\theorembodyfont{\rmfamily} }
\newcommand{\bke}[1]{\left ( #1 \right )}

\newcommand{\bket}[1]{\left \{ #1 \right \}}
\newcommand{\norm}[1]{\left \| #1 \right \|}
\newcommand{\ran}{\rangle}
\newcommand{\lan}{\langle}

\newcommand{\EN}{{L^\infty_t H^1}}
\newcommand{\ST}{{L^2_tW^{1,6}}}
\newcommand{\SD}{{L^2_tW^{1,6/5}}}

\newcommand{\weak}[1]{{\text{w-}#1}}
\newcommand{\lec}{{\ \lesssim \ }}
\newcommand{\gec}{{\ \gtrsim \ }}
\newcommand{\EQAL}[1]{\begin{equation} \begin{split} #1
 \end{split} \end{equation}}
\newcommand{\EQ}[1]{\begin{equation} #1 \end{equation}}

\newcommand{\sidenote}[1]{}
\newcommand{\LR}[1]{{\lan #1 \ran}}
\newcommand{\C}{\Complex}
\newcommand{\fy}{\varphi}
\newcommand{\de}{\delta}
\newcommand{\si}{\sigma}
\newcommand{\imply}{{\quad\text{implies that}\quad}}

\begin{document}

\title{Asymptotic Stability and Completeness in the Energy Space for
Nonlinear Schr\"odinger Equations with Small Solitary Waves}

\author{Stephen Gustafson, \quad Kenji Nakanishi, \quad Tai-Peng Tsai}

\date{gnt12.tex,\quad 2003-08-06}

\maketitle

\section{Introduction}

In this paper we study a class of nonlinear Schr\"odinger equations
which admit families of small solitary wave solutions. 
We consider solutions which are small in the energy space $H^1$,
and decompose them into solitary wave and dispersive wave components.  
The goal is to establish the asymptotic stability of the solitary wave 
and the asymptotic completeness of the dispersive wave.
That is, we show that as $t \rightarrow \infty$, 
the solitary wave component converges to a fixed solitary wave,
and the dispersive component converges to a solution of the
free Schr\"odinger equation.  

Let us briefly supply some background.  Solutions of dispersive
partial differential equations (with repulsive nonlinearities) tend to
spread out in space, although they often have conserved $L^2$
mass. There has been extensive study of this phenomenon, usually
referred to as {\it scattering theory}. These equations include
Schr\"odinger equations, wave equations, and KdV equations.  However,
such equations can also possess solitary wave solutions which have
localized spatial profiles that are constant in time (e.g., if the
nonlinearity is attractive or if a linear potential is present).  To
understand the asymptotic dynamics of general solutions, it is
essential to study the interaction between the solitary waves and the
dispersive waves.  The matter becomes more involved when the
linearized operator around the solitary wave possesses multiple
eigenvalues, which correspond to excited states. The interaction
between eigenstates (mediated by the nonlinearity) is very delicate,
and few results are known.

For nonlinear Schr\"odinger equations with solitary waves, there are three
types of results:

1. Control of solutions in a finite time interval, and construction
of all-time solutions with specified asymptotic behaviors 
({\it scattering solutions}, see \cite{FTY,FG}).  
This type of result does not allow sufficient time for the 
excited state interaction to make a difference.

2. Orbital stability of solitary waves. A solution stays close to the
family of nonlinear bound states if it is initially close.  This is
usually proved by energy arguments, see e.g. \cite{CL,W2,GSS,RW}.

3. Asymptotic stability of solitary waves. Here, one must 
assume that the spectrum of the
linearized operator enjoys certain spectral properties 
(for example, has only
one eigenvalue, or has multiple ``well-placed'' eigenvalues).  
Furthermore, {\it the initial data are typically
assumed to be localized}, so that the dispersive wave has fast local
decay. Even under restrictive spectral assumptions, only
perturbation problems can be treated for large solitary waves (see
\cite{C1,C2}, also \cite{BP1,BP2,BS} for 1-D results), while
more general results can be obtained for small solitary waves
\cite{SW1,PW,Wd,TY1,TY2,TY3,TY4,T}.

In this paper, we study small solutions of the equation
\begin{equation}\label{NLS}
i \pd_t \psi = (- \Delta + V)\psi + g(\psi), \qquad
\psi(0,\cdot) = \psi_0 \in H^1(\R^3)
\end{equation}
with small data: $\|\psi_0\|_{H^1} << 1$
(this is equivalent to considering a nonlinearity multiplied by
a small constant).  Although we
only consider the problem for $x\in \R^3$, the results and methods can
be extended to spatial dimensions $d \ge 3$.

Here, $g(\psi)$ is either a pointwise nonlinearity or a Hartree-type
(non-local) nonlinearity (or their sum), satisfying gauge covariance:
\EQ{
\label{gcov}
  g(\psi e^{i\al})= g(\psi)e^{i\al},
  \quad \mbox{ with } \quad g(|\psi|) \in \R.
} 
More detailed assumptions are given below.
In either case, we can find a functional
$G: H^1\to\R$, satisfying 
$G(\psi e^{i\al})=G(\psi)$
(gauge invariance), 
and
\EQ{
 \pd_\e^0 G(\psi+\e\eta) := 
  \frac{d}{d\e} G(\psi+\e\eta) |_{\e = 0} 
 =\Re(g(\psi),\eta).
}
Here we denote the inner product in $L^2$ by
\EQ{
 (a,b) := \int_{\R^3} \bar{a}b \, dx.
} 
Under suitable assumptions, the $L^2$-norm $\norm{\psi(t)}_{L^2}$ 
and the Hamiltonian 
\begin{equation}
 \frac{1}{2} \int_{\R^3} 
  \bke{|\nabla\psi|^2 + V|\psi|^2 }dx + G(\psi)  
\end{equation}
are constant in time. Using these conserved quantities, and the
smallness of $\norm{\psi_0}_{H^1}$, one can prove a uniform estimate
$\sup_t \norm{\psi(t)}_{H^1} \ll 1$ and obtain global
well-posedness. This is, however, not used in our proof.

We do {\it not} assume that $\psi_0$ is localized ($\psi_0 \in
L^1(\R^3)$, for example, or in a weighted space), as is usually done.
As a result, we cannot expect a time decay rate for $L^p$-norms ($p >
2$) of the dispersive part of the solution.  However, the space $H^1$
is natural, as it is intimately related to the Hamiltonian structure,
and persists globally in time (in contrast to weighted spaces, whose
smallness persists only for short time due to dispersion, and $L^1$,
which may be instantaneously lost and so does not seem to have
physical relevance).  A related motivation comes from the situation
where the linearized operator around a solitary wave has many
eigenvalues.  In this case, the dispersive component tends to decay
very slowly.  It is thus essential to be able to remove the
localization assumption on the data.

We assume that $- \Delta + V$ supports only one eigenvalue $e_0<0$,
which is non-degenerate, and we denote by $\phi_0$ the corresponding
positive, normalized eigenfunction.  More detailed assumptions on $V$
are given below.  Under these assumptions, there exists a family of
small ``nonlinear bound states'' $Q=Q[z]$, parameterized by small
$z=(\phi_0,Q)\in\C$, which satisfy $Q[z] - z \phi_0 = o(z) \perp
\phi_0$, and solve the nonlinear eigenvalue problem
\begin{equation} \label{Q.eq}
 (- \Delta + V)Q + g(Q) = E Q , \qquad 
 E = E[z] = e_0 + o(z) \in\R.
\end{equation}
See Lemma \ref{lemma1} for details. 
Gauge covariance is inherited by $Q$:
\EQ{
 Q[ze^{i\al}] = Q[z] e^{i\al},
}
and so $E[z]=E[|z|]$. 
The nonlinear bound states give rise to exact 
solitary-wave solutions 
$\psi(x,t) = Q(x)e^{-iEt}$ of \eqref{NLS}.
$Q[z]$ is differentiable in $z$ if we regard it as a real vector 
\EQ{
 z=z_1 + i z_2 \leftrightarrow (z_1,z_2) \in \R^2.
}
We will denote its $z$-derivatives by
\EQ{
 D_1Q[z] := \frac{\pd}{\pd z_1} Q[z], \quad 
 D_2Q[z] := \frac{\pd}{\pd z_2} Q[z]
}
(we use the symbol $D$ in order to distinguish them from 
space or time derivatives). 
Then $DQ[z]$ denotes the Jacobian matrix,
regarded as a $\R$-linear map on $\C$:
\EQ{
 DQ[z]:\C\to\C,\quad DQ[z]w \mapsto D_1Q[z]\Re w + iD_2Q[z]\Im w.
}
The gauge covariance of $Q[z]$ implies that
\EQ{ \label{gauge:DQ}
 DQ[z]iz = iQ[z].
}

Given a general solution $\psi(t)$ of~\eqref{NLS}, it is natural to
decompose it into solitary wave and dispersive wave
components:
\EQ{ \label{split}
  \psi(t) = Q[z(t)] + \eta(t).
}
For any such decomposition, \eqref{NLS} yields an equation for $\eta$:
\EQ{ \label{eq:eta}
 i\dot\eta = H[z]\eta + E[z]Q[z] -iDQ[z]\dot z + F_2(z,\eta),
}
where $H[z]$ denotes the {\it linearized operator} around $Q[z]$, 
\EQ{ \label{H.def}
  H[z]\eta := (-\Delta+V)\eta + \pd_\e^0 g(Q+\e\eta),
}
and $F_2$ collects terms which are higher-order in $\eta$: 
\EQ{ \label{F2.def}
  F_2(z,\eta) := g(Q+\eta)-g(Q)-\pd_\e^0 g(Q+\e\eta).
}

The decomposition~\eqref{split} is of course not unique.
To specify the path $z(t)$ uniquely, we impose an orthogonality
condition which will make $\eta$ dispersive.
Since the linearization destroys gauge invariance,
the linearized operator $H[z]$ is not complex-linear.
It is, however, symmetric if we regard  
$\C$ as $\R^2$, and use the reduced inner product:
\EQ{ 
 \LR{a,b} := \Re(a,b) = 
 \int_{\R^3} \bke{ \Re a \Re b + \Im a \Im b } \, dx.
}
The symmetry of $H[z]$ follows from \eqref{H.def} and
\EQ{
 \LR{\pd_\e^0 g(Q+\e\eta),\zeta}
  =\pd_\e^0 \LR{g(Q+\e\eta),\zeta}
  =\pd_\e^0 \pd_\de^0 G(Q+\e\eta+\de\zeta). 
}
We define the ``continuous spectral subspace'' 
\begin{equation} \label{Hcz.def}
  \Hc[z] := \bket{\eta \in L^2: 
  \LR{i\eta,D_1Q[z]}=\LR{i\eta,D_2Q[z]}=0}.
\end{equation} 
This is an invariant subspace of $i(H[z] - E[z])$,
as follows from the relation
\EQ{
\label{inv}
  (H[z]-E[z]) D_j Q[z] = (D_j E[z]) Q[z]
}
(which is the result of differentiating~\eqref{Q.eq}),
together with~\eqref{gauge:DQ}.
Restricting to $\Hc[z]$ eliminates 
non-decaying solutions of the linear equation
$\pd_t \eta = -i(H[z]-E[z]) \eta$ for fixed $z$.
When $z\in \R^+$, $\Hc[z]$ is just the orthogonal complement of 
$\{Q,i \frac \pd{\pd |z|}Q\}$ in the inner product 
$\lan \cdot, \cdot \ran$
(this subspace is often encountered in the literature).

As we will show in Lemma \ref{lemma4}, we can uniquely decompose 
$\psi(t)$ as
\begin{equation} \label{psi.dec}
  \psi (t)= Q[z(t)] + \eta (t), \qquad \eta(t) \in \Hc[z(t)].
\end{equation}
The requirement $\eta(t) \in \Hc[z(t)]$ determines $z(t)$ uniquely.
An evolution equation for $z(t)$ is derived from differentiating the
relation $\LR{i\eta,D_jQ[z]}=0$ with respect to $t$, and using
equation \eqref{eq:eta} (see \eqref{z.eq}).  Our goal is to prove the
asymptotic stability of $Q[z(t)]$ and the asymptotic completeness of
$\eta(t)$.

We now state precise assumptions on the potential $V$, and on the
nonlinearity $g$.  We denote the usual Lorentz space by
$L^{p,q}=(L^\infty,L^1)_{1/p,q}$ for $1<p<\infty$ and $1\le q
\le\infty$ (see \cite{BL}).  $W^{1,p}$ denotes the usual Sobolev
space.

\medskip

{\bf Assumption 1}: 
$V$ is a real-valued function belonging to $L^2 + L^\infty$.
(We note that under this assumption, $-\Delta + V$ is a
self-adjoint operator on $L^2$, with domain $H^2$.
See, eg. \cite{RS}).
Its negative part $V_- := \max \{ 0, -V \}$ is
further assumed to satisfy 
$\norm{V_-}_{(L^2 + L^\infty)(\{|x|>R\})} \to 0$ as $R \to \infty$.
We suppose $-\Delta + V$ has only one eigenvalue $e_0<0$,
and let $\phi_0$ be a corresponding
normalized eigenvector. $e_0$ is simple and $\phi_0$ can be taken to 
be positive (\cite{RS}). Denote the projections onto the
discrete and continuous spectral subspaces 
of $-\Delta + V$ by
\begin{equation} \label{Pc.def}
  P_d = \phi_0(\phi_0,\cdot), \qquad  P_c = 1- P_d.
\end{equation}
The following Strichartz estimates are assumed to hold:
\EQAL{ \label{Strz}
&\norm{e^{it(\Delta-V)}P_c\phi}_{X} \lec \|\phi\|_{H^1},\\
&\norm{\int_{-\infty}^t e^{is (\Delta - V)} P_c F(s)ds }_{X} 
\lec \norm{F}_{\SD},
}
where $X := \EN \cap \ST \cap L^2_t L^{6,2}$.

We remark that the Strichartz estimates of
Assumption 1 hold when, e.g., 
\begin{equation} \label{ass:1b}
 |V(x)| \le C (1+|x|)^{-3-\e},
\end{equation}
for some $\e>0$, and the bottom of the continuous spectrum, zero, is
neither an eigenvalue nor a resonance.  Estimates without derivatives
can be proved by applying the $L^1$-$L^\infty$ decay estimate
\cite{JSS,Y,GS} to the endpoint Strichartz estimate \cite{KT}, where
the stronger estimate in the Lorentz space was actually proved. We
need the Lorentz space $L^{6,2}$ estimate only to handle the critical
case of the Hartree equation (with convolution potentials decaying
like $1/|x|^2$).
Estimates of the derivatives can be obtained by using the equivalence
\sidenote{change C}
\begin{equation} \label{eq124}
  \norm{ \phi}_{W^{1,p}} \sim \| H^{1/2} \phi \|_{L^p} 
  \qquad H=-\Delta+V+\norm{V}_\infty+1
\end{equation}
for $1<p<\infty$, and the commutativity of $e^{it (\Delta-V)}$ with
$H^{1/2}$.  The equivalence can be shown by applying the complex
interpolation for fractional powers \cite[\S 1.15.3]{Tri} to the
equivalence in $W^{2,p}$, using the boundedness of imaginary powers
$H^{is}$, which follows by \cite{Co} from the fact \sidenote{ref. is
Cowling now} that $e^{-tH}$ is a positivity preserving contraction
semigroup on $L^p$.

{\bf Assumption 2}: The nonlinearity is assumed to be 
of one of the following two forms, or their sum:

(a) $g : \C \to \C$ is a function satisfying
gauge covariance \eqref{gcov}
which, when restricted to $\R$, is twice differentiable,
with $g(0)=g'(0)=0$, and  
\EQ{ \label{ass:2a}
  |g''(s)|\le C (s^{1/3} + s^3).
}

(b) $g(\psi) = (\Phi * |\psi|^2)\psi$, 
where $\Phi$ is a real potential, and
\EQ{ \label{ass:2b} 
  \Phi \in L^1 + L^{3/2,\infty}.
}

Examples of nonlinearities satisfying Assumption 2 include
\sidenote{power changed}
\begin{equation}
  g(\psi) = a|\psi|^{4/3} \psi + b |\psi|^4 \psi 
  + \big [\big (\frac{c}{|x|^{3-\e}} + \frac{d}{|x|^{2}} \big )* 
  |\psi|^2 \big ] \psi ,  
\end{equation}
where $a,b,c,d \in \R$, $0<\e < 1$.

We can now state our main theorems. 

\begin{theorem}[Asymptotic stability and completeness]
\label{thm1}
Let Assumptions 1 and 2 hold.
Every solution $\psi$ of \eqref{NLS} with data 
$\psi_0$ sufficiently small in $H^1$
can be uniquely decomposed as 
\EQ{
 \psi(t)= Q[z(t)] + \eta(t),
}
with differentiable $z(t) \in \Complex$ and $\eta(t)\in \Hc[z(t)]$
satisfying
\EQAL{
 \norm{\eta}_{\ST \cap \EN} 
 + \norm{z}_{L^\infty_t} \lec \norm{\psi_0}_{H^1},\\
 \norm{\dot z + iE[z]z}_{L^1_t \cap L^\infty_t} \lec \norm{\psi_0}_{H^1}^2.
}
Moreover, there exist $m_\infty \ge 0$ with 
$\big|m_\infty-|z(0)|\big| \lec \|\psi_0\|_{H^1}^2$,
and $\eta_+ \in H^1\cap \mr{Ran}\, P_c$ such that
\EQ{ \label{asymptotic}
 |z(t)|\to m_\infty, \quad 
 \norm{\eta(t)-e^{it(\Delta-V)}\eta_+}_{H^1} \to 0 
}
as $t \to \infty$.
\end{theorem}

The corresponding result with no bound state was obtained in 
\cite{S,JSS} for small $H^1$ data and \cite{GV} for large data with 
no potential and $g(\psi)=+ |\psi|^{m-1}\psi$.
Results similar to Theorem \ref{thm1} in the case of {\it localized}
initial data $\psi_0$ and $g(\psi)=\lambda |\psi|^{m-1}\psi$ were
first obtained for the case $\norm{\eta(0)}_{H^1 \cap L^1} \ll |z(0)|$
by Soffer and Weinstein \cite{SW1}, and then extended to all $\psi_0$
small in $H^1$ and weighted $L^2$-spaces by Pillet and Wayne
\cite{PW}.  
The latter work was extended to the 1-D case in \cite{Wd}.
In all \cite{SW1,PW,Wd}, the solutions $\psi(t)$ are
decomposed with respect to fixed self-adjoint linear operators. A
time-dependent decomposition similar to \eqref{psi.dec} seems to have
first appeared in \cite{BP1}.

\begin{theorem}[Nonlinear wave operator]\label{thm2}
Let Assumptions 1 and 2 hold. There exists $\de>0$ such that for any
$m_\infty \in [0,\de]$ and $\eta_+ \in H^1\cap \mr{Ran}\, P_c$ with
$\norm{\eta_+}_{H^1} \le \de$, there is a global solution $\psi(t)$ of
\eqref{NLS} satisfying the conclusion of Theorem~\ref{thm1} with the
prescribed asymptotic data $m_\infty$ and $\eta_+$.
\end{theorem}

Special cases of Theorem \ref{thm2}, further assuming $m_\infty \gg
\norm{\eta_+}_{H^1 \cap L^1}$ or $m_\infty=0$ with $\norm{\eta_+}_{H^1
\cap L^1}\ll 1$ for $g(\psi)=\pm |\psi|^2\psi$, were obtained in
\cite{TY1,TY3}.

In Theorems \ref{thm1} and \ref{thm2}, one may replace
$e^{it(\Delta-V)}\eta_+$ by $e^{it\Delta}\widetilde \eta_+$ with
$\widetilde \eta_+ \in H^1$ if asymptotic completeness
in $H^1$ of the wave operator between $-\Delta+V$ and $-\Delta$ holds. 
It holds, for example, if \eqref{ass:1b} holds.

Theorem \ref{thm1} implies in particular that any small solution looks
like a solitary wave for large time locally in space. But 
due to the fact that the data is not assumed to be localized,
we cannot, in general, have a convergence rate for it. 
In fact we have

\begin{theorem}[Examples of slow decay of dispersion] \label{thm3}
Let Assumptions 1 and 2 hold. For any nonempty ball $B\subset \R^3$,
there exists $\de>0$ for which the following holds.  For any positive
function $f(t)$ which goes to zero as $t \to \infty$, and any
$m_\infty\in[0,\de]$, there exists a solution $\psi(t)$ of \eqref{NLS}
satisfying the conclusions of Theorem~\ref{thm1}, and 
\EQ{
  \limsup_{t\to\infty} \bke{ \inf_{|z'|\le 2\de} 
\norm{\psi(t)-Q[z']}_{L^2(B)}/f(t) }  =\infty.  
}
\end{theorem}

\noindent
{\bf Remarks:}
Note that we impose a time-dependent condition $\eta(t) \in \Hc[z(t)]$
instead of simpler conditions such as $\eta(t) \in \Hc[0]$, i.e.,
$(\eta(t),\phi_0)=0$ (which is used in \cite{PW,Wd}).
The reason is the following. 
If we assume $(\eta(t),\phi_0)=0$, then the
equations for $\dot z+iEz$ yield
\[
  |\dot z + iEz|\lec |(\phi_0, A\eta)| + |(\phi_0, F_2)|,
\]
where $A$ is some linear operator.  The term $(\phi_0, A\eta)$ 
is linear in $\eta$ and hence is not integrable in time, in
light of the estimate $\eta \in L^2_t W^{1,6}$. Thus we cannot
conclude that $|z|$ and $E[z]$ have limits as $t \to \infty$.  This
term drops out if we require $\eta(t) \in \Hc[z]$, and the equation
for $\dot z + iEz$ (and hence $\frac d{dt} |z|$) becomes quadratic in
$\eta$. Even without this term, we are forced to use an 
$L^2_t$-type Strichartz estimate
in order to get convergence of $|z|$, since we cannot have better
decay as long as we start with $H^1$ initial data.  Thus the end point
Strichartz estimates \eqref{Strz} are irreplaceable in our argument. As
a bonus, they allow us to treat borderline nonlinearities such as 
$g(\psi)=\pm |\psi|^4 \psi$ and $g(\psi)=\pm |\psi|^{4/3} \psi$,
which are not covered in the previous works \cite{
SW1,PW}.
By considering the problem in higher Sobolev spaces $H^s$, $s>1$,
one may
treat higher power nonlinearities $g(\psi)=\pm |\psi|^{m-1} \psi$ for
$m\in (5, 1+4/(3-2s))$.


\section{Preliminaries}

In this section, we give three lemmas concerning the nonlinear
ground states. 

\begin{lemma}[Nonlinear ground states] \label{lemma1}
There exists $\delta>0$ such that 
for each $z \in \C$ with $|z|\le \delta$, there is a solution 
$Q[z] \in H^2 \cap W^{1,1}$ of \eqref{Q.eq} with $E=E[|z|]\in\R$
such that
\[
  Q[z]=z \phi_0 + q[z] , \qquad (q,\phi_0)=0.
\]
The pair $(q,zE)$ is unique in the class
\[
  \norm{q}_{H^2} \le \delta, \qquad |E- e_0| \le \delta.
\]
Moreover,
$Q[ze^{i\al}]=Q[z]e^{i\al}$, $Q[|z|]$ is real, and
\EQAL{ 
 &\left.
  \begin{split} 
   &q[z]=o(z^2)\\ 
   &DQ[z] = (1,i)\phi_0 + o(z),\quad 
   D^2Q[z] = o(1)
  \end{split}
  \right\}\quad\text{in $H^2 \cap W^{1,1}$}, \\
 &E[z] = e_0 + o(z),\quad DE[z]=o(1),
}
as $z\to 0$.  
\end{lemma}

A special case of this lemma is proved in~\cite{SW1},
referring to~\cite{Ag}.
We will prove the lemma under weaker assumptions on $V$
and $g$ in the Appendix.

The following is an immediate but useful corollary of this lemma. 
\begin{lemma}[Continuous spectral subspace comparison]
\label{lemma3}
There exists $\de>0$ such that, for each $z\in\C$ with $|z|\le\de$, 
there is a bijective operator $R[z]:\Hc[0]\to\Hc[z]$ satisfying  
\EQ{
 P_c|_{\Hc[z]} = R[z]^{-1}. 
}
Moreover, $R[z]-I$ is compact and continuous in $z$ in the operator norm 
on any space $Y$ satisfying 
$H^2\cap W^{1,1}\subset Y\subset H^{-2}+L^\infty$.  \end{lemma}
We remark that no corresponding statement holds for the 
case of large solitary waves. 
\myproof
$R[z]$ is given by
\EQ{
 R[z]\eta = \eta + \phi_0\al[z]\eta,
}
where the operator $\al[z]:\Hc[0]\to\C$ is defined by solving the equations 
\EQ{
 \LR{\eta+\phi_0\al[z]\eta,D_j Q[z]} = 0, \quad j=1,2.
}
This is solvable due to the property $D_j Q[z]=(1,i)\phi_0+o(z)$. 
Then $R[z]$ is obviously the inverse of $P_c$ restricted onto $\Hc[z]$. 
For any $\eta\in\Hc[0]$ we have
\EQ{
 |\al[z]\eta| \lec |\LR{\eta,DQ[z]}| \lec o(z)\|\eta\|_{H^{-2}+L^\infty},
}
which implies compactness of $R[z]-I$ in $Y$. The continuity in $z$ follows 
from that of $DQ[z]$. 
\QED

\begin{lemma}[Best decomposition] 
\label{lemma4}
There exists $\de>0$ such that any $\psi\in H^1$ satisfying 
$\norm{\psi}_{H^1} \le \de$ can be uniquely decomposed as 
\begin{equation} 
\label{eq1:lemma4}
  \psi = Q[z] + \eta, \
\end{equation}
where $z \in \Complex$, $\eta \in \Hc[z]$ and 
\(
 |z|+\norm{\eta}_{H^1} \lec \|\psi\|_{H^1}. 
\)
\end{lemma}

\myproof
We look for a unique solution $z$ of the equation $A(z)=0$,
where we define   
\[
  A_j(z) := \LR{i(\psi - Q[z]),D_jQ[z]}, \qquad j=1,2.
\]
Let $n:=\norm{\psi}_{H^1}$. The Jacobian matrix of the map 
$z\mapsto A(z)$ is written as 
\EQAL{ \label{DA}
  D_jA_k(z) &=
  \LR{-iD_jQ[z],D_kQ[z]} +\LR{i(\psi-Q[z]),D_jD_kQ[z]}\\
  &= j-k + o(n+|z|),
}
by Lemma \ref{lemma1}. Let $z_0:=(\phi_0,\psi)$.
So $|z_0| \le n$.  Then from Lemma \ref{lemma1} we have
\EQ{
 A(z_0) = \LR{i(\psi-z_0\phi_0)+o(n^2),(1,i)\phi_0+o(n)}=o(n^2). 
}
Now the result is an immediate consequence of the inverse function theorem.
\QED


\section{Asymptotic stability and completeness}

This section is devoted to a proof of Theorem~\ref{thm1}.  
We will first estimate the nonlinearity in Subsection 3.1, 
and then prove the theorem in the subsequent subsection.

\subsection{Estimates on the nonlinearity}
Before starting the proof of our theorems, 
we establish some nonlinear estimates, first 
for the pointwise nonlinearity, and then for the 
convolution nonlinearity.  

\medskip
\noindent {\bf (I) Pointwise nonlinearity}

Our assumption \eqref{ass:2a} implies that for $k=0,1,2$ 
\EQ{
 |D^kg(z)| \lec \sum_{j=0}^k |g^{(j)}(|z|)z^{j-k}| \lec |z|^{7/3-k}+|z|^{5-k}.
}
The nonlinear term $F_2$, defined in \eqref{F2.def}, 
can be expanded by the mean value theorem as
\EQ{ 
\label{decp F1}
  F_2(\eta)=g(Q+\eta)-g(Q)-\pd_\e^0g(Q+\e\eta)
  =\int_0^1 (1-\e)\pd_\e^2 g(Q+\e\eta) d\e.
}
Then we can estimate it as 
\[
 |F_2| \lec \sup_{0<\e<1} |D^2g(Q+\e\eta)||\eta|^2 
  \lec (1+|Q|+|\eta|)^4|\eta|^2,
\]
\EQ{ 
\label{eq35}
 \norm{F_2}_{L^1+L^\infty} \lec (1+\|Q\|_{L^6}+\|\eta\|_{L^6})^4
 \|\eta\|_{L^6}^2.
}
We also need to estimate $g(Q+\eta)-g(Q)$.  
By using the generalized H\"older inequality,
\EQAL{  \label{eq36}
 &\norm{g(Q+\eta)-g(Q)}_{W^{1,6/5}} 
  \lec \norm{(Dg(Q+\eta)-Dg(Q))\nabla Q}_{L^{6/5}} 
  +\norm{Dg(Q+\eta)\nabla\eta}_{L^{6/5}}\\
 &\lec (\|Q\|_{L^2}+\|\eta\|_{L^2})^{1/3}\|\eta\|_{L^6}\|\nabla Q\|_{L^2}
 + (\|Q\|_{L^6}+\|\eta\|_{L^6})^3\|\eta\|_{L^6}\|\nabla Q\|_{L^6}\\
 & \qquad \qquad + (\|Q\|_{L^2}^{4/3}+\|\eta\|_{L^2}^{4/3}+\|Q\|_{L^6}^4
   +\|\eta\|_{L^6}^4)\|\nabla\eta\|_{L^6}\\
 &\lec C(\|Q\|_{H^2}+\|\eta\|_{H^1})\|\eta\|_{W^{1,6}},
}
where $C(s)\lec s^{4/3}+s^4.$

\medskip
\noindent {\bf (II) Convolution nonlinearity}

The nonlinear term $F_2$ has the following form:
\[ 
 F_2 = Q \Phi*|\eta|^2 + \eta \Phi*(2\Re(Q\bar{\eta}) + |\eta|^2).
\]
By the generalized Young inequality in Lorentz spaces, 
we have under the assumption \eqref{ass:2b},
\begin{equation} \label{eq37}
  \norm{F_2}_{L^1+L^\infty} 
  \lec (\norm{Q}_{L^6}+\norm{\eta}_{L^6})\norm{\eta}_{L^{6,2}}^2.
\end{equation}
This is the only place where we need the Lorentz space $L^{6,2}$. 

As for $g(Q+\eta)-g(Q)$, its gradient $\nabla(g(Q+\eta)-g(Q))$ 
is expanded into a sum of trilinear forms where one of three functions 
has the derivative and at least one of them is $\eta$. 
By the generalized Young inequality, we have
\EQ{
 \|\Phi*(\psi_1\psi_2)\psi_3\|_{L^{6/5}}
 \lec \|\Phi\|_{L^1+L^{3/2,\infty}} \|\psi_{\si(1)}\|_{L^2}
 \|\psi_{\si(2)}\|_{L^2 \cap L^6}\|\psi_{\si(3)}\|_{L^6}, 
}
for any permutation $\si$. So we may put an $\eta$ or $\nabla\eta$ in $L^6$, 
another function without derivative in $L^2\cap L^6$,
and the remaining one in $L^2$. Hence we obtain
\EQ{ \label{eq39}
 \|g(Q+\eta)-g(Q)\|_{W^{1,6/5}} 
 \lec \|\eta\|_{W^{1,6}}(\|\eta\|_{H^1}+\|Q\|_{H^1})^2.
}

\subsection{Asymptotic stability and completeness}
\label{sec:stab}

Now we prove our main result, Theorem \ref{thm1}. 

Let $\psi(x,t)$ solve the
nonlinear Schr\"odinger equation~(\ref{NLS})
with initial data
\[
  \psi(0,\cdot) = \psi_0, \qquad \norm{\psi_0}_{H^1(\R^3)} \ll 1.
\]

It is easy to prove local well-posedness in $H^1$ by using the 
Strichartz estimate \eqref{Strz} (the discrete spectral part does 
not bother us on finite intervals). 
The unique solution thereby obtained belongs to $\EN \cap \ST$. 

Our argument below will yield time-global a priori estimates, 
so that the solution $\psi$ exists and remains small in $H^1$ for all time. 
More precisely, we take $\de'>0$ much smaller than any $\de$ in the 
previous lemmas, and take the initial data $\psi_0$ such that 
\EQ{
  \|\psi_0\|_{H^1} < \de' \ll \de. 
}
We will show that 
\EQ{ \label{bootstrap}
 \|\psi\|_{\EN[0,T]}< \de \imply \|\psi\|_{\EN[0,T]} < \de/2,
}
for any $T>0$, provided $\de$ and $\de'$ were chosen sufficiently small. 
Then, by continuity in time, this bound and the solution together extend 
globally in time. In the argument below we will not explicitly specify the 
time interval. The assumption $\|\psi\|_{\EN}<\de$ allows us to use 
all of the previous lemmas. 

By Lemma~\ref{lemma4} we have the decomposition
\[
  \psi = Q[z(t)] + \eta(t), \qquad
  \eta \in \Hc[z].
\]
The equation for $\eta$ is given in \eqref{eq:eta}.  We now derive the
evolution equation for $z(t)$. Differentiating the relation
$\LR{i\eta,D_jQ[z]}=0$ with respect to $t$ and 
plugging equation~\eqref{eq:eta} into that, we obtain
\[
  0=\LR{H\eta + EQ -iDQ\dot z +F_2,D_jQ}+\LR{i\eta,D_jDQ\dot z},
\] 
where $H$, $E$ and $Q$ all depend on $z$ (but this dependence is
dropped from the notation). 
By the symmetry of $H$ and~\eqref{inv}, we have
\EQ{
 \LR{H\eta,D_jQ} = \LR{\eta,HD_jQ}= \LR{\eta,D_j(EQ)}
 =\LR{\eta,ED_jQ}=\LR{i\eta,E D_j DQ iz},
}
where we used $\LR{i\eta,DQ}=0$. By \eqref{gauge:DQ}, we have
\EQ{
 \LR{EQ-iDQ\dot z,D_jQ}= \LR{DQ(iEz+\dot z),iD_jQ}.
}
Thus we obtain
\EQ{ \label{z.eq}
 \sum_{k=1,2}(\LR{iD_jQ,D_kQ}+\LR{i\eta,D_jD_kQ})(\dot z+iEz)_k 
 = -\LR{F_2,D_jQ}. 
}
The matrix on the left hand side is the Jacobian matrix in \eqref{DA}, 
and so is estimated as
\EQ{
 \LR{iD_jQ,D_kQ}+\LR{i\eta,D_jD_kQ}=j-k + o(\de).
}
Inverting this matrix, we obtain 
\EQ{ \label{param est}
  |\dot{z}+iEz| \lec  |\LR{F_2, DQ[z]}| \lec \|F_2\|_{L^1 + L^\infty}
}
at any $t$. Applying the estimates \eqref{eq35} and \eqref{eq37}, we obtain
\EQ{ \label{param est2}
  \|\dot z+iEz\|_{L^2} \lec 
  \|\eta\|_{L^\infty_t L^{6,2}}\|\eta\|_{L^2_t L^{6,2}} 
  \lec \|\eta\|_{\EN}\|\eta\|_{L^2_t L^{6,2}},
}
and $\|\dot z+iEz\|_{L^\infty}\lec \norm{\eta}_{L^\infty_t H^1}^2$,
where we used the Sobolev embedding $H^1\subset L^{6,2}$. 

Next, we estimate $\eta$ by writing the equation \eqref{eq:eta}
in the form 
\begin{equation}
\label{pde2}
  i \pd_t \eta = (-\Delta + V)\eta + F
\end{equation}
with
\begin{equation} 
\label{def F}
  F := g(Q+\eta)-g(Q)  - iDQ(\dot{z}+iEz).  
\end{equation}
Denote $\eta_c := P_c \eta$ where
$P_c = 1-\phi_0(\phi_0,\cdot)$ is defined in \eqref{Pc.def}.
The Strichartz estimates applied to~(\ref{pde2}) and 
Lemma \ref{lemma3} yield
\EQ{ \label{eta est}
  \|\eta\|_{X} \lec \|\eta_c\|_{X} \lec \|\eta(0)\|_{H^1}
  + \| P_c F \|_{\SD} \lec \|\psi_0\|_{H^1} + \|F\|_{\SD},
}
where 
\EQ{
 X:= \EN \cap \ST \cap L^2_t L^{6,2}.
}
By Lemma~\ref{lemma1} and the estimates \eqref{eq36} and \eqref{eq39} 
in the previous subsection, we obtain
\EQ{ \label{pde est}
  \|F\|_{\SD} \lec    \| \dot{z}+iEz \|_{L^2}   +\de \| \eta \|_{\ST}.
}
From \eqref{param est2}, \eqref{eta est} and \eqref{pde est}, we deduce that 
\EQ{
  \| \eta \|_{X} + \| \dot{z}+iEz \|_{L^2}^{1/2} +\|F\|_{\SD} 
  \lec \|\psi_0\|_{H^1}<\de',
}
if we take $\de$ sufficiently small. Choosing $\de'$ even smaller, 
we obtain the desired bootstrapping estimate \eqref{bootstrap}, 
and so the solution, as well as all the estimates, extends globally.

Moreover, we have
\EQ{
 \|\pd_t|z|\|_{L^1} \le \|\dot z + iEz\|_{L^1} \lec 
 \|\eta\|_{L^2_t L^{6,2}}^2 \lec \|\psi_0\|_{H^1}^2,
}
so $|z(t)|$ and $E[z(t)]=E[|z(t)|]$ converge as $t\to\infty$.

Finally, we prove that $\eta$ is asymptotically free. 
We have the integral equation 
\begin{equation} \label{int eq scatt}
  \eta_c(t) = e^{it(\Delta-V)} \left[\eta_c(0) -i \int_0^t
  e^{-is(\Delta-V)} P_c F(s) ds \right].
\end{equation}
By the Strichartz estimate, for any $T>S>0$ we have
\[
 \norm{\int_S^T e^{-is(\Delta-V)} P_c F(s) ds}_{H^1} 
 \lec \norm{F}_{\SD[S,T]}\to 0,
\]
as $T>S\to\infty$, by the Lebesgue dominated convergence theorem, and
the finiteness of $\norm{F}_{\SD(0,\infty)}$. Thus the integral in
\eqref{int eq scatt} converges in $H^1$ as $t\to\infty$, and we obtain
\[
  \lim_{t\to\infty} e^{-it(\Delta-V)}\eta_c(t)=\eta_c(0) -i \int_0^\infty
  e^{-is(\Delta-V)} P_c F(s) ds=:\eta_+.
\]
In particular, $\eta_c(t)$ converges to $0$ weakly in $H^1$.  Then
Lemma \ref{lemma3} implies that $\eta_d(t)=(R[z(t)]-I)\eta_c(t)$
converges to $0$ strongly in $H^1$.  Therefore we conclude that
\[
  \norm{\eta(t)-e^{it(\Delta-V)}\eta_+}_{H^1} \to 0.
\]
\QED

 
\section{Nonlinear wave operator}

In this section we prove Theorem \ref{thm2}.
We will construct the desired solution by first assigning the
asymptotic data at large finite time $T$ and then taking the weak
limit as $T\to\infty$.  
Recall that $m_\infty, \norm{\eta_+}_{H^1}\le \de$.
For any $T>0$, we define $\psi^T$
to be the solution of \eqref{NLS} with the initial condition
\[ 
  \psi^T(T)=Q[m_\infty] + e^{iT(\Delta-V)}\eta_+.
\]
Theorem \ref{thm1} implies that $\psi^T$ is globally defined 
uniquely and satisfies
\EQ{ \label{bds}
 \norm{\psi^T}_{\ST \cap L^\infty H^1} \lec \de,
   \quad \norm{\pd_t|z^T|}_{L^1(T,\infty)} 
   \le \norm{(\dot z + iEz)^T}_{L^1(T,\infty)} \lec \de^2, 
}
where the solution can be decomposed according to 
Lemma~\ref{lemma4} as 
\[ 
  \psi^T = Q[z^T] + \eta^T, \quad \eta^T\in \Hc[z^T].
\]
Let $\xi_T:=e^{-iT(\Delta-V)}P_c \eta^T(T)$. 
Then we have the integral equation for any $T>0$
\[
  P_c \eta^T(t) = e^{it(\Delta-V)}\xi_T 
  -i \int_T^t e^{i(t-s)(\Delta-V)}P_c F^T(s) ds,
\]
where $F^T$ is as defined in \eqref{def F}. 
By the same argument as in Subsection \ref{sec:stab}, 
we deduce that for any $S\le T$,
\EQAL{
 &\norm{\eta^T}_{\ST[S,\infty]} \lec \norm{P_c \eta^T}_{\ST[S,\infty]} 
 \lec \norm{e^{it(\Delta-V)}\xi_T}_{\ST[S,\infty]} 
 + \norm{F^T}_{\SD[S,\infty]}\\
 &\norm{F^T}_{\SD[S,\infty]} \lec \norm{(\dot z+iEz)^T}_{L^2[S,\infty]} 
 + \de\norm{\eta^T}_{\ST[S,\infty]}
 \lec \de\norm{\eta^T}_{\ST[S,\infty]}. 
}
Therefore, when $\de$ is sufficiently small, we have 
\[ 
  \norm{\eta^T}_{\ST[S,\infty]} \lec 
  \norm{e^{it(\Delta-V)}\xi_T}_{\ST[S,\infty]}.
\]
Applying the Strichartz estimate once again, we get
\begin{equation} 
\label{diff from scat}
  \norm{P_c \eta^T - e^{it(\Delta-V)}\xi_T}_{\EN\cap\ST\cap 
  L^2_t L^{6,2}[S,\infty]} \lec \de 
  \norm{e^{it(\Delta-V)}\xi_T}_{\ST[S,\infty]}.
\end{equation}

Now we take the limit $T\to\infty$. 
First we check the convergence of the data at $t=T$.
Denote $z_T:=z^T(T)$ and $\eta_T:=\eta^T(T)$. Since $z_T$ is bounded, 
it converges to some $z_\infty\in\C$ along some subsequence. Then we have
\EQAL{
 0 &= \LR{i\eta_T, DQ[z_T]i(m_\infty-z_\infty)}\\
 &=\LR{i(Q[m_\infty]-Q[z_\infty]), DQ[z_\infty]i(m_\infty-z_\infty)}
 = |m_\infty-z_\infty|^2(1-O(\de)),
}
where the second equality is by taking limits, and we used 
Lemma~\ref{lemma1}, that 
$\eta_T = Q[m_\infty]-Q[z_T]-e^{iT(\Delta-V)}\eta_+$ and that 
$e^{iT(\Delta-V)}\eta_+$ converges to $0$
weakly in $H^1$. Thus 
\EQ{
 z^T(T) \to m_\infty, \quad \xi_T \to \eta_+ \text{ in } H^1,
}
as $T\to\infty$ (without restriction to a subsequence). 

Now we proceed to convergence for all time.  By \eqref{bds}, $z^T$ is
equicontinuous on $\R$, and so is $|z^T|$ on the extended real line
$[-\infty,\infty]$.  From the equations, $\eta^T$ is equicontinuous in
$C(\R,\weak{H^1})$, so is $P_c\eta^T$ in
$C([-\infty,\infty],\weak{H^1})$ by \eqref{diff from scat}.  Then by
Lemma \ref{lemma3}, $\eta^T=R[z^T]P_c\eta^T$ is also equicontinuous in
$C([-\infty,\infty],\weak{H^1})$.  Therefore $\eta^T$ and $z^T$ are
convergent along some subsequence in the following topology:
\EQAL{
 &\eta^T\to\eta^\infty, \quad \text{in } (C^0\cap L^\infty)(\weak{H^1})\cap \weak{\ST},\\
 & z^T\to z^\infty, \quad \text{in } C^0(\R),
 \quad |z^T| \to |z^\infty| \quad \text{in } L^\infty(\R).
}
This implies the convergence of $\psi^T$ itself:
\begin{equation} \label{scatt limit}
 \psi^T= Q[z^T] + \eta^T \to Q[z^\infty] + \eta^\infty =: \psi^\infty,
\end{equation}
in $C(\weak{H^1})\cap\weak{\ST}$ on any finite time interval. 
Extracting a subsequence if necessary, we may assume that the nonlinearity 
$g(\psi^T)$ also converges in $\weak{\SD}$ on any finite time interval. 
Then the local convergence of $\psi^T$ in $C(L^p)$ for $p<6$ implies that 
the limit of the nonlinearity is the desired $g(\psi^\infty)$. 
Hence we deduce that $\psi^\infty$ is a solution to \eqref{NLS} belonging to 
$C(\R;H^1)\cap L^2_{loc}(W^{1,6})$. 
From the uniform convergence of $|z^T|$ to $|z^\infty|$, and 
the convergence of $z^T(T)$ we have
\[
  \lim_{T \to \infty} |z^\infty(T)| = \lim_{T\to\infty}|z^T(T)| = m_\infty.
\]
From convergence of $\xi_T$, \eqref{diff from scat} and the 
weak convergence uniform in time, we get
\EQAL{
 &\|P_d\eta^\infty(t)\|_{H^1}\to 0,\quad \text{as $t\to\infty$}\\
 &\norm{P_c\eta^\infty - e^{it(\Delta-V)}\eta_+}_{\EN[S,\infty]} 
 \lec\norm{e^{i(\Delta-V)t}\eta_+}_{\ST[S,\infty]} \to 0,\quad \text{as }
S\to\infty.
}
Thus $\psi^\infty$ is a solution with the desired asymptotic profile. 
\QED


\section{Examples of slow decay of dispersion}

In this section we prove Theorem \ref{thm3}.

For a fixed ball $B \subset \R^3$,
choose $\xi_0\in H^1$ satisfying
\EQ{
 \|\xi_0\|_{H^1}=1, \quad \|\xi_0\|_{L^2(B)}>0, \quad 
 (\phi_0,\xi_0)=(\phi_0,\xi_0)_{L^2(B)}=0,
}
where the first inner product is in $L^2(\R^3)$ as before.  The
constants below depend on $B$ and $\xi_0$. We use a small parameter
$\e>0$ to control the size of solution.

We define an increasing  sequence of times $T_j$ inductively as follows. 
Let $T_j>1$ and for $j>1$, assume that we have defined $T_k$ for $k<j$. 
By the Strichartz estimate, there exists $T>\max_{k<j}T_k$ such that 
\EQ{ \label{Strz bd} 
 \sum_{k<j,\ \pm} \|e^{\pm i(\Delta-V)(t-T_k)}\xi_0\|_{\ST[T,\infty]} 
 < \e 2^{-j}
}
and 
\EQ{ \label{ft bd}
 \sup_{t>T} f(t) \le \e^2 2^{-2j}. 
}
Then we can choose $T_j>T$ such that 
\EQ{ \label{L6 bd}
 \sum_{k<j,\ \pm} \|e^{\pm i(\Delta-V)(T_j-T_k)}\xi_0\|_{W^{1,6}} < \e 2^{-j}.
}
We define the final data by 
\[ 
  \eta_+ := \sum_{j>0} \e 2^{-j} e^{i(\Delta-V)(-T_j)} \xi_0,
\]
and the asymptotic profile of the dispersive part is given by
\EQ{
  \eta_\ell := \sum_j \e 2^{-j} e^{i(\Delta-V)(t-T_j)} \xi_0.
}

Let $\psi(t) = Q[z(t)] + \eta(t)$ be the solution furnished
by Theorem~\ref{thm2} corresponding to $\eta_+$ and $m_\infty \in [0,\e]$.
By \eqref{diff from scat} and \eqref{Strz bd} we have
\EQ{
 \|P_c \eta(T_j) - \eta_\ell(T_j)\|_{H^1} \lec \e\|\eta_\ell\|_{\ST[T_j,\infty]}
 \lec \e^2 2^{-j}. 
}
By \eqref{L6 bd} we have
\EQ{
 \|\eta_\ell(T_j) - \e 2^{-j}\xi_0\|_{L^6} \lec \e^2 2^{-j}. 
}
Then by Lemma \ref{lemma3} we have
\EQ{
 \|P_d \eta(T_j)\|_{L^6} \lec \e\|P_c\eta(T_j)\|_{L^6} \lec \e^2 2^{-j}.
}
Thus we obtain
\EQ{ \label{aprx eta j}
 \|\eta(T_j) - \e 2^{-j}\xi_0\|_{L^2(B)} \lec \|\eta(T_j) - \e 2^{-j}\xi_0\|_{L^6} \lec \e^2 2^{-j}. 
}
By Lemma~\ref{lemma1}, we have
\[
  \|Q[z(T_j)]-Q[z] - (z(T_j)-z)\phi_0\|_{L^2(B)} \lec o(\e)|z(T_j)-z|.
\]
Since $\phi_0>0$ everywhere, we may assume, by choosing $\e$ sufficiently small, that
\EQ{ \label{diff Q}
  \|Q[z(T_j)]-Q[z] - (z(T_j)-z)\phi_0\|_{L^2(B)} < \e \|(z(T_j)-z)\phi_0\|_{L^2(B)}. 
}
Since $\xi_0$ and $\phi_0$ are orthogonal in $L^2(B)$, we obtain from \eqref{aprx eta j} and \eqref{diff Q}, 
\[
  \| \eta(T_j) + Q[z(T_j)] - Q[z] \|_{L^2(B)} 
  \gec \e 2^{-j}\|\xi_0\|_{L^2(B)} - \e^2 2^{-j} \gec \e 2^{-j}, 
\]
provided $\e$ is sufficiently small. 
Thus we obtain by \eqref{ft bd} 
\[
  \inf_z \norm{ \psi(T_j) - Q[z]}_{L^2(B)} \gec \e 2^{-j} \gec 2^{j} f(T_j). 
\]
\QED


\section{Appendix: nonlinear bound states}
\newcommand{\N}{{\cal N}}
\newcommand{\B}{{\cal B}}
\newcommand{\T}{T}
\newcommand{\E}{\EuScript{E}}

In this appendix we prove Lemma~\ref{lemma1}.

For the linear potential $V$, we may weaken Assumption 1 to include
only those parts which are relevant for existence of nonlinear bound
states:

\smallskip
\noindent {\bf Assumption 1$'$}: We suppose $V \in L^2 + L^\infty$
with $\| V_- \|_{L^2 + L^\infty}(|x| > R) \to 0$ as $R \to \infty$,
and that $e_0 < 0$ is a simple eigenvalue of $-\Delta + V$ (we do not
need the Strichartz estimates, and we do not need $e_0$ to be the only
eigenvalue, or even the ground state).
\smallskip
 
We need the nonlinearity $g$ to be just super-quadratic. Thus we may
replace Assumption 2 by the following:

\smallskip
\noindent {\bf Assumption 2$'$:} $g$ is as in Assumption 2, but in the
pointwise case is only required to satisfy (instead of \eqref{ass:2a})
\[
  g''(s) = o(1) \quad \mbox{ as } \quad  s \in \R \to 0^+.
\]
\smallskip

\noindent 
{\bf Proof of Lemma \ref{lemma1} under Assumptions 1$'$ and 2$'$:}
\smallskip

For each $z$, we look for a solution $Q = z \phi_0 + q$ and $E=e_0 +
e'$ of \eqref{Q.eq} with $(\phi_0,q)=0$ small and $e' \in \R$ small.
Then
\[
  (-\Delta + V)q + g(z\phi_0 + q) = e' (z\phi_0 + q) + e_0 q.
\]
Taking projections on the $\phi_0$ and $\phi_0^\perp$ directions, we
get
\begin{gather}\label{eq1:L21}
  e' z = (\phi_0, g(z\phi_0 + q) ) \\
  \label{eq2:L21}
  \T q = - P_c g(z\phi_0 + q) + e'q,
\end{gather}
where we denote $\T:=-\Delta+V-e_0$. 
The right sides are of order $o(z)$. 
We will use a contraction mapping argument to solve for $q=o(z^2)$ in
$H^2$ and $e'=o(z)$ uniquely, for sufficiently small $z$.
Differentiating by $z$, we obtain the equations for higher derivatives:
\EQAL{
 &zDe' + e'J = (\phi_0, Dg(Q)),\\
 &zD^2e' + J D e' = (\phi_0, D^2 g(Q)),\\
 &\T Dq = -P_cDg(Q) + q De' + e'Dq,\\
 &\T D^2q = -P_c D^2g(Q) + q D^2e' + Dq De' + e'D^2q,\\
 &Dg(Q) = g'(Q)(J\phi_0+Dq),\\
 &D^2g(Q) = g''(Q)(J\phi_0+Dq)^2 + g'(Q)D^2q,
}
where we have omitted subscripts for $D$ and 
$J := Dz = (1,i)$, 
and some constant coefficients. 

Assumption 1$'$ implies that 
$-\Delta + V$ is self-adjoint on $L^2$ with
domain $H^2$, so $\phi_0\in H^2$.  
Furthermore, the assumption
$\|V_-\|_{(L^2+L^\infty)(\{|x|>R\})} \to 0$
implies that $-\Delta + V$ is a relatively compact 
perturbation of $-\Delta + V_+$.  So by the Weyl theorem,
the essential spectrum of $-\Delta + V$ is contained in 
$[0,\infty)$.  In particular,
$e_0$ is an isolated point of the spectrum.
Since it is a simple eigenvalue, we have 
\EQ{
  \T^{-1}: L^2_\perp \to H^2_\perp \text{ bounded},
}
where $H^{2}_\perp$ and $L^2_\perp$ denote the Sobolev spaces 
restricted to the orthogonal complement of $\phi_0$. 

Now we can solve the equations in the closed convex set
\EQ{
  K:= \{(q,e') \in H^2_\perp \times \R \mid 
  \|q\|_{H^2} \le |z|^2, |e'| \le |z|\}
}
for sufficiently small $z\in\C$ (the case $z=0$ is trivial). 
Indeed, we define the map 
$M:(q_0,e'_0) \mapsto (q_1,e'_1)$ by
\EQAL{
 &g_0 := g(z\phi_0+q_0),\\
 &ze'_1 := (\phi_0, g_0),\\
 &q_1 := \T^{-1}(-P_c g_0 + e'_0q_0).
}
Then under Assumption 2$'$ we have the easy estimates
\EQAL{
  &|ze'_1| \lec \|g_0\|_{H^{-2}} \le o(\|z\phi_0+q_0\|_{H^2}^2) \lec o(z^2)\\
  &\|q_1\|_{H^2} \lec \|g_0\|_{L^2} + |e'_0|\|q_0\|_{H^2} \lec o(z^2),
}
which implies that $M$ maps $K$ into $K$. 
Let $(e'_{j+2},q_{j+2}):=M(e'_j,q_j)$ and $g_j:=g(z\phi_0+q_j)$  for $j=0,1$. 
Similarly, we can estimate the difference 
\EQAL{
  &|z(e'_2-e'_3)| \lec \|g_0-g_1\|_{H^{-2}} \le o(\|z\phi_0+q_j\|_{H^2})
  \|q_0-q_1\|_{H^2}
  \lec o(z)\|q_0-q_1\|_{H^2},\\
  &\|q_2-q_3\|_{H^2} \lec \|g_0-g_1\|_{L^2} + |e'_0-e'_1|\|q_0\|_{L^2} 
  + |e'_1|\|q_0-q_1\|_{H^1}\\
  & \qquad \lec |z|(|e'_0-e'_1|+\|q_0-q_1\|_{H^2})
}
which implies that $M$ is a contraction on $K$, 
and hence has a unique fixed point in $K$. 

Suppose now there is a solution $(q,e')$ in the class
$K'=\bket{(q,e'):\norm{q}_{H^2} \le \gamma, |e'|
\le \gamma}$. We have
\[
  \norm{q}_{H^2} \lec \norm{g(z\phi_0 + q)}_{L^2} + |e'| \norm{q}_{L^2}
  \lec o(1) |z|^2 + o(1) \norm{q}_{H^2}.
\]
Thus $\norm{q}_{H^2} \lec o(1) |z|^2$. It follows that
$|ze'| \lec  \norm{g(z\phi_0 + q)}_{H^{-2}} \lec o(z^2)$ and hence 
$(q,e')\in K$. This shows the uniqueness in the class $K'$.

Let $(q,e')$ be the unique solution and $Q:=z\phi_0+q$. Since the equation 
becomes real-valued when $z\in\R$, the unique solution $Q[z]$ is also 
real-valued. 

By the same argument as above, we have
\EQAL{
 &|z||De'| \lec o(z) + \|Dg(Q)\|_{H^{-2}},\\
 &\|Dq\|_{H^{2}} \lec \|Dg(Q)\|_{L^2} + |z|^2|De'|+|z|\|Dq\|_{H^2},\\ 
 &\|Dg(Q)\|_{L^2} \le o(\|Q\|_{H^2})\|J\phi_0+Dq\|_{H^2} 
  \le o(z)(1+\|Dq\|_{H^2}),
}
which imply that $De'=o(1)$ and $\|Dq\|_{H^2}=o(z)$. 
Similarly we have
\EQAL{
 &|z||D^2e'| \lec o(1)+\|D^2g(Q)\|_{H^{-2}},\\
 &\|D^2q\|_{H^{2}} \lec \|D^2g(Q)\|_{L^2} + o(z^2)|D^2e'| 
 + o(z)|De'|+ o(z)\|D^2q\|_{H^2},\\
 &\|D^2 g(Q)\|_{L^2} \lec o(1)\|J\phi_0+Dq\|_{H^2}^2 + o(z)\|D^2q\|_{H^2},
}
which imply that $|D^2e'|=o(1/z)$ and $\|D^2q\|_{H^2}=o(1)$.

Next we establish the estimate in $W^{1,1}$. 
Actually, following~\cite{Ag}, we will obtain an 
exponentially weighted energy estimate. Let
\[
  \E(\psi,\phi) := (\nabla \psi,\nabla \phi) 
  + \int V \bar{\psi} \phi dx
\]
denote the bilinear form associated to $-\Delta + V$.
It is defined on $H^1 \times H^1$.  Set
\EQ{
 b := \lim_{R\to\infty} \inf \{\E(\fy,\fy) \mid
 \fy \in H^1, \|\fy\|_2=1, 
 \fy(x) = 0 \text{ for $|x| < R $} \}.
}
Suppose $b<0$. Then there exists a sequence $\fy_R$ satisfying 
$\|\fy_R\|_2=1$, $\fy_R(x)=0$ for $|x|<R$, and $\E(\fy_R,\fy_R)<\de$ 
for some fixed $\de<0$.  It is easy to check that 
$\fy_R$ is bounded in $H^1$. Since it converges weakly to $0$ as $R\to\infty$,
by the assumption $\|V_-\|_{(L^2+L^\infty)(\{|x| > R\})} \to 0$, 
the negative part $\int V_- |\fy_R|^2 dx$ of the energy converges
to $0$, a contradiction. Thus $b \ge 0$. 
In other words, there exists $\de(R)$ with $\de(R) \to b \geq 0$ 
as $R\to\infty$, such that for any $\fy\in H^1$ satisfying 
$\fy(x)=0$ for $|x|<R$, we have
\EQ{ \label{far energy}
 \E(\fy,\fy) \ge \de(R)\|\fy\|_2^2. 
}
Now we apply this inequality with localized exponential weight 
$\chi_R$ defined by
\EQ{
 \chi_R(x) = 
  \begin{cases}
   e^{\e(|x|-R)}-1,&(R<|x|<2R),\\
   e^{\e(3R-|x|)}-1,&(2R<|x|<3R),\\
   0 \text{ else}
  \end{cases}
}
with some fixed small $\e>0$. Then we have
\EQ{ \label{diff exp} 
 |\nabla \chi_R| \lec \e(\chi_R+1).
}
From \eqref{far energy}, we have for any $\fy\in H^1$, 
\EQAL{
 \de(R)\|\chi_R\fy\|_2^2 \le \E(\chi_R\fy,\chi_R\fy) 
 &= \E(\chi_R^2\fy,\fy)+\int |\fy\nabla\chi_R|^2 dx\\
 &=(\chi_R^2\fy,\T\fy)+e_0\|\chi_R\fy\|_2^2 + \int |\fy\nabla\chi_R|^2 dx.
}
By using \eqref{diff exp}, we obtain for sufficiently large $R$ and small $\e$,
\EQ{
 \|\chi_R\fy\|_2^2 \lec (\chi_R^2\fy,\T\fy) + \e^2\|\fy\|_2^2.
}
The Sobolev norm also is estimated by 
\EQAL{
 \|\chi_R\nabla\fy\|_2^2 &\le \|\nabla(\chi_R\fy)\|_2^2 
 + \|\fy\nabla\chi_R\|_2^2\\
 &\lec \E(\chi_R\fy,\chi_R\fy)+ \|\chi_R\fy\|_2^2 + \e^2\|\fy\|_2^2\\
 &\lec (\chi_R^2\fy,\T\fy) + \e^2\|\fy\|_2^2.
}
Thus we obtain the key relation
\EQ{
 \label{key}
 \|\chi_R\fy\|_{H^1}^2 \lec (\chi_R\fy,\chi_R\T\fy) + \|\fy\|_2^2,
}
for any $\fy\in H^1$, sufficiently large $R$ and small $\e$. 
Now we substitute each of $\fy=\phi_0,q,Dq,D^2q$ into~\eqref{key}
and use the equations they satisfy. We find
\EQ{
 \|\chi_R\phi_0\|_{H^1}^2 \lec \|\phi_0\|_2^2 =1,
}
and under Assumption 2$'$ we easily obtain
\EQAL{
 \|\chi_Rq\|_{H^1}^2 &\lec \|Q^{-1}g(Q)\|_{{\cal B}(H^1;H^{-1})}
(\|\chi_RQ\|_{H^1}\|\chi_Rq\|_{H^1} 
 + \|Q\|_{H^1}\|\chi_Rq\|_{H^1}\|\chi_R\phi_0\|_{H^1})+\|q\|_2^2\\
 &\lec (o(z^2)+|z|\|\chi_Rq\|_{H^1})\|\chi_Rq\|_{H^1} + o(z^4), 
}
which implies that $\|\chi_Rq\|_{H^1}=o(z^2)$.
Similar estimates hold for $\chi_RDq$ and $\chi_RD^2q$. 
It follows that each of these functions are bounded in $W^{1,1}$.

\section*{Acknowledgments}
Part of the work was done when all of us were visiting the Banff
International Research Station (BIRS). We wish to thank BIRS and its
sponsors MSRI, PIMS, and the Banff Center, for providing us with a
very stimulating and fruitful research environment.  The research of
the first and third authors is partly supported by NSERC grants
nos. 22R80976 and 22R81253.  The second author is partly supported by
a JSPS Postdoctoral Fellowship for Research Abroad.

\noindent{Stephen Gustafson},  gustaf@math.ubc.ca \\
Dept. Mathematics, University of British Columbia, 
Vancouver, BC V6T 1Z2, Canada

\bigskip

\noindent{Kenji Nakanishi}, knakanis@math.princeton.edu \\
Dept. Mathematics, Princeton University, Princeton,
NJ 08540, USA

\bigskip

\noindent{Tai-Peng Tsai},  ttsai@math.ubc.ca \\
Dept. Mathematics, University of British Columbia, 
Vancouver, BC V6T 1Z2, Canada

\end{document}